\documentclass[reprint, superscriptaddress, secnumarabic, amssymb, nobibnotes, aps, prl]{revtex4-1}

\usepackage{graphicx}
\usepackage{epstopdf}
\usepackage[T1]{fontenc}
\usepackage[latin9]{inputenc}
\usepackage{amsbsy}
\usepackage{gensymb}
\usepackage[T1]{fontenc}
\usepackage[latin9]{inputenc}
\usepackage{amsmath}
\usepackage{amssymb}
\usepackage{bbm}
\usepackage{physics}
\usepackage{braket}
\usepackage{xcolor}
\allowdisplaybreaks
\usepackage{graphicx}
\usepackage[colorlinks=true]{hyperref}  
\hypersetup{
    bookmarks=true,         
    unicode=false,          
    pdftoolbar=true,        
    pdfmenubar=true,        
    pdffitwindow=false,     
    pdfstartview={FitH},    
    pdftitle={Superconducting properties of pseudobinary telluride Chevrel Phase Mo$_4$Re$_2$Te$_8$ },    
    pdfauthor={, , ,},     
    pdfsubject={},   
    pdfcreator={},   
    pdfproducer={}, 
    pdfkeywords={} {} {}, 
    pdfnewwindow=true,      
    colorlinks=true,       
    linkcolor=blue, 
    citecolor=blue,        
    filecolor=magenta,      
    urlcolor=blue           
} 
\usepackage[normalem]{ulem}

\newcommand{\equref}[1]{Eq.~(\ref{#1})}

\newcommand{\figref}[1]{Fig.~\ref{#1}}

\begin{document}
\title{\textrm{Superconducting properties of pseudobinary telluride Chevrel Phase Mo$_4$Re$_2$Te$_8$}}
\author{A. Kataria}
\affiliation{Department of Physics, Indian Institute of Science Education and Research Bhopal, Bhopal, 462066, India}
\author{T. Agarwal}
\affiliation{Department of Physics, Indian Institute of Science Education and Research Bhopal, Bhopal, 462066, India}
\author{S. Sharma}
\affiliation{Department of Physics, Indian Institute of Science Education and Research Bhopal, Bhopal, 462066, India}
\author{A. Ali}
\affiliation{Department of Physics, Indian Institute of Science Education and Research Bhopal, Bhopal, 462066, India}
\author{R. S. Singh}
\affiliation{Department of Physics, Indian Institute of Science Education and Research Bhopal, Bhopal, 462066, India}
\author{R. P. Singh}
\email[]{rpsingh@iiserb.ac.in}
\affiliation{Department of Physics, Indian Institute of Science Education and Research Bhopal, Bhopal, 462066, India}
\date{\today}

\begin{abstract}
Unconventional superconductivity in the Chevrel phase offers a wide structural aspect to understand the superconducting ground state. A detailed investigation on the superconducting properties of Re based pseudobinary telluride Chevrel phase Mo$_4$Re$_2$Te$_8$ is reported. It crystallizes in a trigonal structure with the space group $R\bar{3}H$ having superconducting transition temperature at $T_C$ = 3.26(3) K. Specific heat measurements suggests a fully gapped superconducting state; however, the proximity of upper critical field value from the Pauli limiting field can be attributed to unconventional nature.
\end{abstract}

\maketitle

\section{Introduction}

Unconventional superconductors are an integral part of a new phase of quantum matter exhibiting additional broken symmetries (e.g. time reversal or rotation symmetry) with the global gauge symmetry, U(1). Chevrel phase (CP) is one of the crucial members of the unconventional superconductor family, providing a broader aspect of interactions and structural impact on the superconducting ground state \cite{Uemura}. This phase has a perplexing array of unconventional superconducting properties such as extremely high upper critical magnetic field, multicomponent and multiband superconductivity, magnetic field driven or magnetically ordered superconductivity, reentrant superconductivity, and many more \cite{hhc,mbsc,mfisc,2gsc,nsc,Hom6x8,REmo6X8}. Additionally, the coupling manner of building blocks (Mo clusters) of classical CPs multi-folds various crystal structures \cite{6strt,7strt2,cmo,scb1} producing quasi-1D superconductor \cite{Q1D} and exhibiting interesting topological properties \cite{Tsc}. 

The CPs are extensively known for their relatively high value of the upper critical field, and it is believed that the high spin-orbit splitting by possibly large Sommerfeld coefficient, $\gamma_n$ ascribed to low-lying phonons responsible for it \cite{bC1,mc}. The observed large superconducting gap value in CPs is similar to the superconductivity mediated by soft phonons (low-lying phonons) as recently observed in IrGe \cite{IG}, BaPd$_2$As$_2$ \cite{BP2A2} and SrPt$_3$P \cite{Sp3}. Further, the superconducting properties of CPs depend on the intercluster Mo-Mo distance \cite{scb1,scb}. The partial substitution at Mo site can change metallic compound Mo$_6$X$_8$ into the semiconducting sulphides and selenides \cite{rurh,mre} and in contrast, tellurides of this family become superconductor \cite{mrute,scb,MRu}. Most studies on the unconventional superconducting properties of CPs have primarily focused on the rare-earth intercalated sulfides and selenides. However, to date, the high upper critical field and pairing mechanism are not understood properly. Along with this, the superconducting properties of Chevrel phase tellurides are largely unexplored. It can provide details about the electron-electron interaction, electron-phonon coupling and spin-orbit coupling effect on the nature of superconductivity and pairing mechanism, which can be helpful to understand the unconventionality in the superconducting state of CPs. Detailed studies of new CPs are essential for deep insight into the superconducting ground state and gap symmetry.\\
In this work, we investigate the superconducting properties of a less explored Re based pseudobinary telluride Chevrel phase Mo$_4$Re$_2$Te$_8$, where Re is partially substituted at the Mo site. This study comprises two different prospects; firstly, the presence of Re in the Mo cluster will increase the spin-orbit coupling strength (SOC $\propto$ Z$^4$) and affect the phononic distribution by affecting the related interactions of the CP system. Secondly, this compound can address the ambiguity in Re-based superconductors around the possible reason for time-reversal symmetry (TRS) breaking. Such as Re$_6$X (X = Zr, Hf, Ti), a noncentrosymmetric family shows spontaneous field presence, breaking TRS, regardless of the element X \cite {re6zr, re6hf, re6ti}, while the other non-centrosymmetric compounds Re$_3$Y (Y = Ta, W) and the Re-B system \cite{re3ta,re3w,reb}, preserve time-reversal symmetry. Furthermore, the uncertainty in TRS breaking presence in centrosymmetric Re is also intriguing \cite{Re,Ren} and raises more questions about the role of associated structure and Re concentration in time reversal symmetry breaking. In this regard, investigating more Re based superconductors are essential and Re based pseudobinary telluride CP Mo$_4$Re$_2$Te$_8$, provides that platform with a new structural aspect and different Re concentration. We have performed the temperature-dependent measurements of AC transport, magnetization, and specific heat in different magnetic fields, which allow deducing the superconducting characteristics parameters with other electronic parameters of Mo$_4$Re$_2$Te$_8$. The extracted upper critical field value is close to the Pauli limiting field, which suggests the possibility of unconventionality in the superconducting ground state. Specific heat measurement indicates a moderate electron-phonon coupling with $s$-wave gap symmetry. Moreover, initial band structure calculations suggests importance of SOC on the electronic states of Mo$_4$Re$_2$Te$_8$ along with dominance of d state of Re and Mo atoms in density of states at the Fermi level.

\section{Experimental Details}

Polycrystalline sample of the nominal composition Mo$_4$Re$_2$Te$_8$ was prepared by solid-state reaction method where the constituent elemental powder of Mo (99.99\%), Re (99.99\%), and Te (99.9999\%) was mixed together in a stoichiometric ratio. The palletized form of the mixture was sealed in an evacuated quartz ampoule and heated between 1150$\degree-$1200$\degree$C for several days. Powder x-ray diffraction (XRD) pattern was collected using a PANalytical diffractometer equipped with Cu $K_{\alpha}$ radiation ($\lambda$ = 1.5406 \text{\AA}). Magnetic measurements were performed on a superconducting quantum interference device of Quantum Design magnetic property measurement system (MPMS 3, Quantum Design). Transport measurements and specific heat were carried out on a physical property measurement system (PPMS). The four-probe technique is used to measure the AC transport, and the two - $\tau$ relaxation method is used for specific heat measurement. 

\section{Results and Discussion}

\subsection{Sample characterization}

\begin{figure} 
\includegraphics[width=1.0\columnwidth, origin=b]{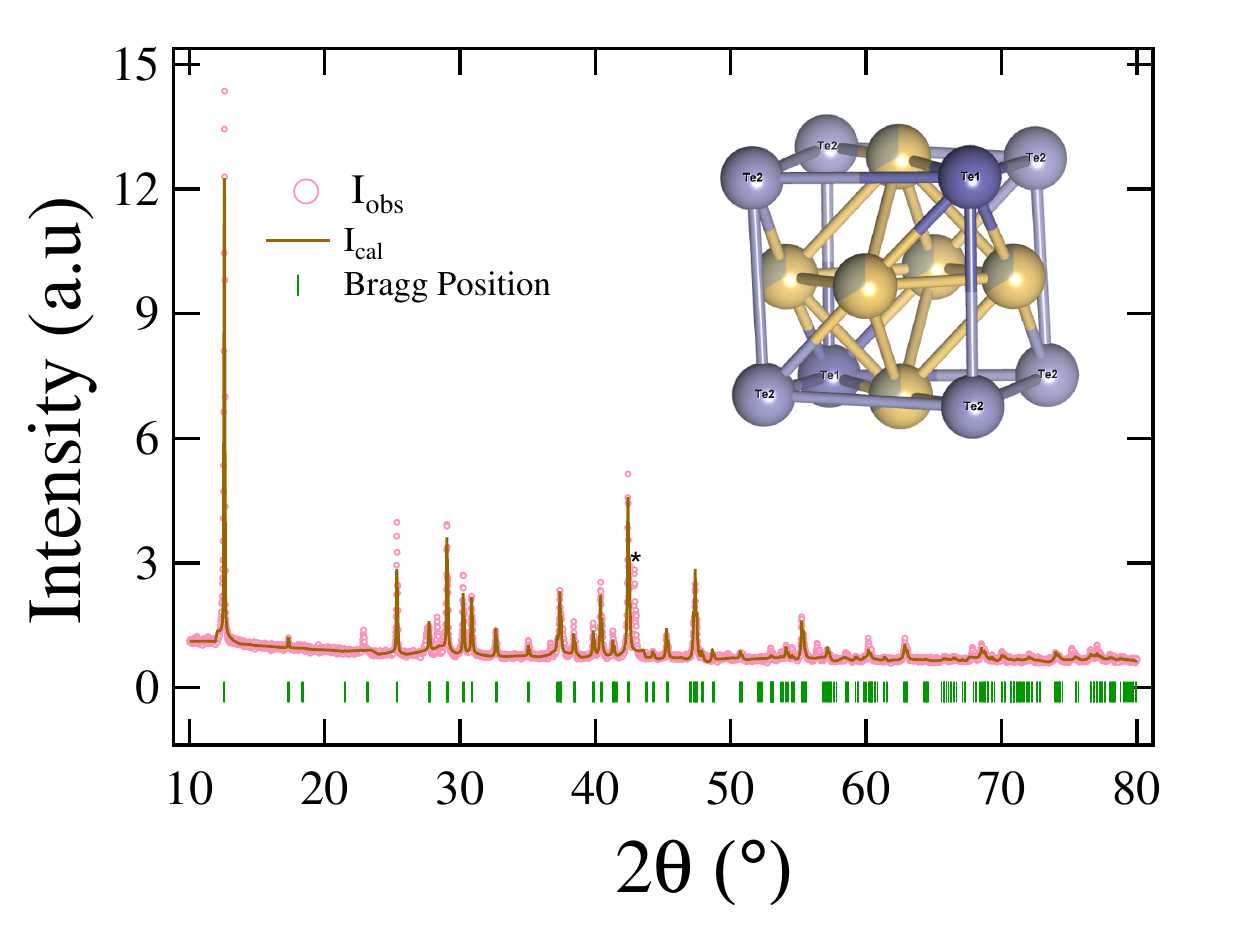}
\caption{\label{fig1:Xrd} Powder x-ray diffraction pattern for Mo$_4$Re$_2$Te$_8$ sample at room temperature. The solid brown line represents the Rietveld refinement of the data, with green vertical bars indicating the Bragg reflection peaks. Inset shows the Mo$_{4}$Re$_2$Te$_8$ cluster.}
\end{figure}

The recorded pattern of XRD for Mo$_4$Re$_2$Te$_8$ was refined using Fullprof software \cite{fp}. Rietveld refinement confirms the trigonal structure having space group $R\bar3H$, and \figref{fig1:Xrd} shows the refined pattern. The refined lattice parameters of Mo$_4$Re$_2$Te$_8$ are tabulated together with the parent compound Mo$_6$Te$_8$ in Table I. Mo$_{6-x}$Re$_x$Te$_8$ ($x$ = 2) cluster, the building block of the crystal structure is displayed in the inset of \figref{fig1:Xrd}.

\begin{table}[h!]
\caption{Structure parameters of Mo$_4$Re$_2$Te$_8$  obtained from the Rietveld refinement of x-ray diffraction.}
\begin{tabular}{l r} \hline
Structure & Trigonal\\
Space group&       $R\bar{3}H$\\ [1ex]
\end{tabular}
\\[1ex]

\begingroup
\setlength{\tabcolsep}{4pt}
\begin{tabular}[b]{c c c c c c}
\hline
 Lattice parameters & Mo$_4$Re$_2$Te$_8$& Mo$_6$Te$_8$ \cite{M6T8} & \\[1ex]
\hline\hline
a = b (\text{\AA}) & 10.233(1) & 10.179(1)\\             
c (\text{\AA}) & 11.530(1) & 11.674(2)\\                       
&$\alpha=\beta=90\degree$,&$\gamma=120\degree$\\
[1ex]
\hline
\end{tabular}
\par\medskip\footnotesize
\endgroup
\end{table}

\subsection{Superconducting and normal state properties}

\subsubsection{Electrical Resistivity}

\begin{figure} [b]
\includegraphics[width=1.0\columnwidth, origin=b]{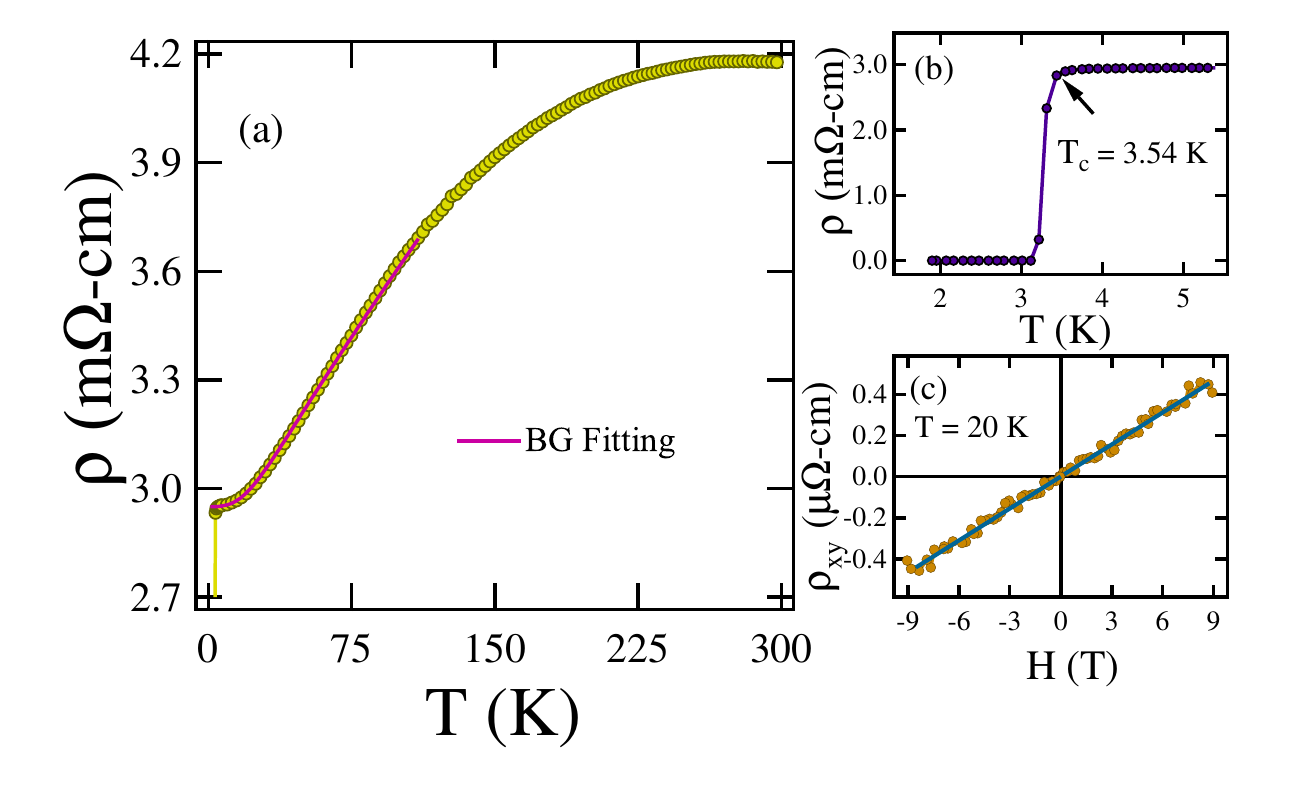}
\caption{\label{fig3:RT}(a) Temperature dependence of resistivity of Mo$_4$Re$_2$Te$_8$ is fitted using BG model shown by the solid purple line. (b) The superconducting drop in resistivity at the transition temperature, $T_C$ = 3.54(2) K. (c) Shows Hall resistivity, $\rho_{xy}$ measurement at 20 K under $\pm$ 9 T magnetic field.}
\end{figure}

\begin{figure*}[ht!] 
\includegraphics[width=2.0\columnwidth, origin=b]{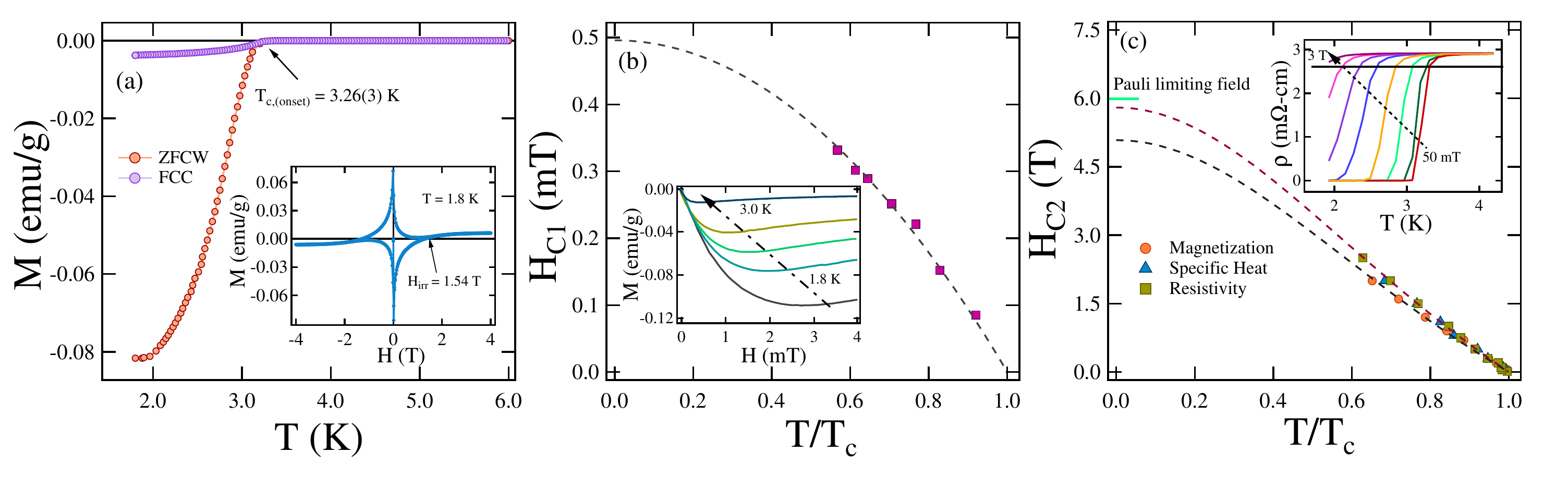}
\caption{\label{fig4:M}(a) The magnetic moment versus temperature for Mo$_4$Re$_2$Te$_8$ with inset showing the magnetization loop at 1.8 K under $\pm$ 4 T magnetic field. (b) Lower critical field estimated from $M$-$H$ curve (inset) at varying temperature. (c) Estimated upper critical field by resistivity, specific heat and magnetization data where dotted lines represent fitting using \equref{eqn4:HC2}. Inset displays resistivity variation with the temperature at different applied fields.}
\end{figure*}

The temperature dependence of resistivity, $\rho(T)$ of the compound Mo$_4$Re$_2$Te$_8$ in zero magnetic field is shown in \figref{fig3:RT}(a). The onset of sharp drop-in resistivity is recorded at temperature $T_{C,onset}$ = 3.54(2) K with the transition width of $\Delta T$ = 0.3 K (\figref{fig3:RT}(b)). The observed $T_C$ matches well with the reported value \cite{mrute,MRu}. The low temperature (in normal state) variation of $\rho (T)$ demonstrates the metallic nature of Mo$_4$Re$_2$Te$_8$ with the increasing behaviour. Hence, the $\rho (T)$ temperature dependence is analysed by using the equation,
\begin{equation}
\rho(T) =  \rho_0 + C\left(\frac{T}{\theta_R}\right)^n \int_{0}^{\theta_R/T}\frac{x^n}{(e^x-1)(1-e^{-x})}dx\\
\label{eqn1:rbg}
\end{equation}

where the second term is the Bloch-Gr\"uneisen expression accounting for the electron-phonon scattering \cite{bg}. $\rho_0$ is the residual resistivity due to defect scattering, $\theta_R$ is debye temperature from resistivity measurements, and $C$ is a material dependent property. $n$ depends on the nature of the interaction. Best fitting of $\rho (T)$ below 100 K, is for $n$ = 3 providing $\theta_R$ =  143(2)K, $C$ = 0.51(1) m$\ohm$-cm and $\rho_0$ = 2.95(1) m$\ohm$-cm. Moreover, the resistivity trend in high temperature region for Mo$_4$Re$_2$Te$_4$ consists of negative curvature ($d\rho/dT$) with the nearly saturated resistivity, indicating an additional contribution in the system with the metallic nature. This behaviour is reminiscent of the other Chevrel phase, skutterudites, and clathrates compounds consisting of similar type of cluster structures \cite{2gsc,cpr,cla,sku,cla1}. Moreover, the same trend is also observed in La$_7$X$_3$ (X = Rh, Ir) family and YCo$_2$, where the electronic scattering is related to the excitations with a broad spectrum of energy is stated the possible reason for the resistivity trend \cite{l7x3,co2}.

Further, the carrier density, $n$, is estimated from the Hall measurement. The longitudinal resistivity, $\rho_{xy}$ measured at 20 K in field dependence, shown in the \figref{fig3:RT}(c). The observed positive slope indicates the dominant hole carriers and provides the Hall resistivity, $R_H$ = $5.2(1) \times 10^{-10} $ m$^{-3}$C. Considering only the single band contribution, the estimated number density $n$ is 1.21(3) $\times$ 10$^{28}$ m$^{-3}$, which is again in good agreement with the known CPs and their selenides substitutes \cite{n,ncp}. 

\subsubsection{Magnetization}

\begin{figure*}[ht!] 
\includegraphics[width=2.0\columnwidth, origin=b]{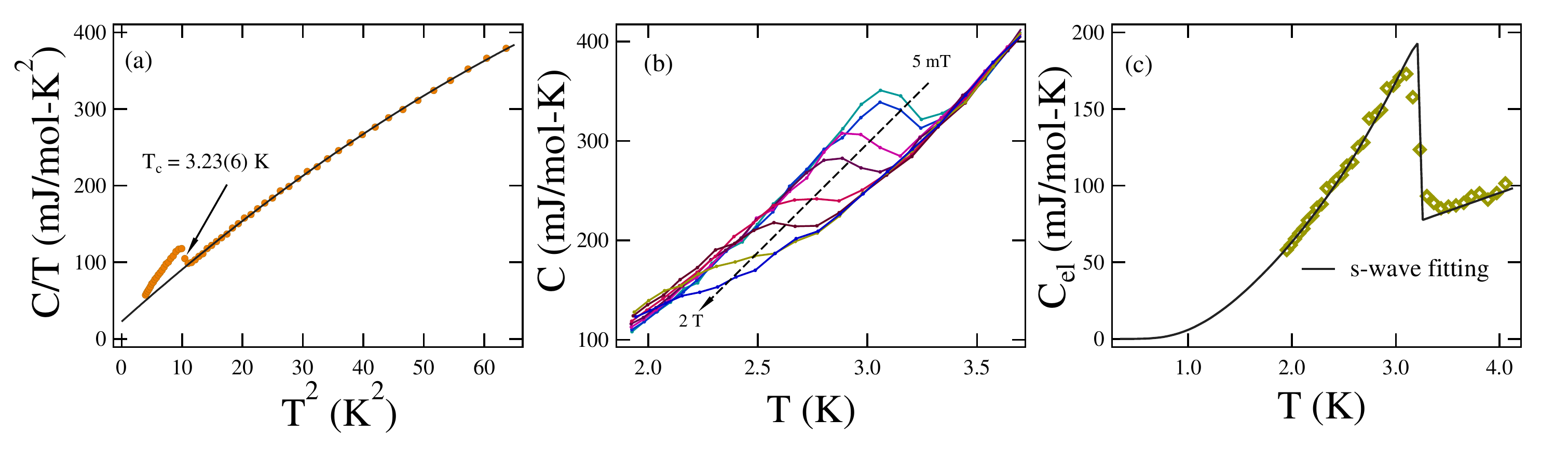}
\caption{\label{Fig6:SH}(a) The low-temperature normal region fitting of $C/T$ vs $T^2$ curve for Mo$_4$Re$_2$Te$_8$ is represented by the solid line. (b) Shows the variation of specific heat jump under magnetic field. (c) Electronic specific heat data, $C_{el}$, is well described by BCS single gap $s$-wave model shown by solid black line. }
\end{figure*} 

The magnetic moment under temperature variation of Mo$_4$Re$_2$Te$_8$ was measured at 1 mT using both zero-field-cooled warming (ZFCW) and field-cooled cooling (FCC) mode, as shown in \figref{fig4:M}(a). The onset of the diamagnetic signal of the compound is noted at $T_{C, onset}$ = 3.26(3) K, where the deviation of FCC curve from the ZFCW depicts the flux pinning in the sample. The inset of \figref{fig4:M}(a) shows the magnetization loop under magnetic field variation $\pm$ 4 T at 1.8 K of Mo$_4$Re$_2$Te$_8$. The full magnetization loop confirms the type-II superconductivity with bulk pinning in the sample representing an irreversible nature of magnetization, $H_{irr}$ = 1.54 T, and a further increase in the applied magnetic field de-pins the vortices.

The lower critical field value, $H_{C1}$(0) is extracted from the magnetization dependence on the magnetic field measured at different temperatures up to $T_C$, as shown in the inset of \figref{fig4:M}(b). The linear deviation from Meissner effect (entering in the vortices region) is considered as the $H_{C1}$ for the respective isotherm. The temperature dependence of $H_{C1}$ for Mo$_4$Re$_2$Te$_8$ is shown in \figref{fig4:M}(b) and fitted by using the Ginzburg-Landau equation given as,

\begin{equation}
H_{C1}(T) = H_{C1}(0)\left[1-\left(\frac{T}{T_{C}}\right)^{2}\right].
\label{eqn3:HC1}
\end{equation} 

The estimated value of $H_{C1}(0)$, is 0.49(1) mT. The upper critical field, $H_{C2}(0)$, value is computed via variation of transition temperature in various applied magnetic fields. The applied, increasing magnetic field shifts the superconducting transition temperature to the lower value as observed in resistivity, magnetization and specific heat measurements. The inset of \figref{fig4:M}(c) indicates the same via resistivity measurement, where $T_C$ is recorded at 90\% drop in normal state resistivity value. The upper critical field value, $H_{C2}$ variation with reduced temperature, $t = T/T_{C}$ was fitted with the Ginzburg-Landau equation,

\begin{equation}
H_{C2}(T) = H_{C2}(0)\left[\frac{(1-t^2)}{(1+t^2)}\right]. 
\label{eqn4:HC2}
\end{equation} 

The well-fitted data is shown in \figref{fig4:M}(c), yielding upper critical field, $H_{C2}(0)$ = 5.08(7) T, 5.62(6) T and 5.78(4) T from magnetization, specific heat and resistivity data, respectively. The effect of an applied external magnetic field is described by two mechanisms involved (1) orbital decoupling and (2) Pauli limiting effect. The orbital limiting field is where the increased kinetic energy of one electron breaks the Cooper pair, given by the Wartherm-Helfand-Hohenberg (WHH) expression \cite{WHH_1,WHH_2},

\begin{equation}
H_{C2}^{orbital}(0) = -\alpha T_{C}\left.\frac{dH_{C2}(T)}{dT}\right|_{T=T_{C}}
\label{eqn5:HHH}
\end{equation}

here $\alpha$ is considered as 0.69 for a dirty limit superconductor. The initial slope $\frac{-dH_{C2}(T)}{dT}$, in the vicinity of $T_C$ is 1.58(1) T for Mo$_4$Re$_2$Te$_8$, which evaluates the orbital limiting field, $H_{C2}^{orbital}(0)$ = 3.55(4) T. From BCS theory of superconductors, another effect Pauli limiting field is represented as $H_{C2}^P$(0) = 1.84 $T_C$ \cite{Pauli_1,Pauli_2}, with $T_C$ = 3.26(3) K, $H_{C2}^P$(0) becomes 5.99(5) T. Moreover, the relative strength of two magnetic effects, orbital and Pauli limiting field, is measured by the Maki parameter \cite{maki}, $\alpha_{M} = \sqrt{2}H_{C2}^{orb}(0)/H_{C2}^{P}(0)$ = 0.84(2). The close value of $\alpha_{M}$ to one indicates the non-negligible effect of Pauli limiting field in the Cooper pair breaking. Along with this, the proximity of upper critical field value ($H_{C2,\rho}(0) = 5.78(4)$ T) with Pauli limiting field is also compelling and suggests the possible unconventional nature associated with the system similar to other Re based superconductors \cite{re6zr,re5ta,HPG}.

Superconducting characteristic length parameters have been calculated from upper and lower critical field values. The relation between $H_{C2}(0)$ and Ginzburg-Landau coherence length, $\xi_{GL}(0)$ \cite{Coh_Leng}, $H_{C2}(0)= \frac{\Phi_0}{2\pi\xi_{GL}(0)^2}$, where $\Phi_0$ = 2.07 $\times 10^{-15}$ Tm$^2$ is the quantum flux, gives $\xi_{GL}(0)$ = 80(1) \text{\AA}. The lower critical field, $H_{C1}(0)$ and coherence length, $\xi_{GL}(0)$ is related to the penetration depth as follows \cite{MIB},

\begin{equation}
H_{C1}(0) = \frac{\Phi_{0}}{4\pi\lambda_{GL}^2(0)}\left(\mathrm{ln}\frac{\lambda_{GL}(0)}{\xi_{GL}(0)}+0.12\right)   
\label{eqn6:PD}
\end{equation} 

providing $\lambda_{GL}(0)$ = 1325(40) nm. The  value of the Ginzburg-Landau parameter, $\kappa_{GL} = \frac{\lambda_{GL}(0)}{\xi_{GL}(0)}$, is calculated to be 165(7), indicating a strong type-II superconductivity in Mo$_4$Re$_2$Te$_8$ consistent with the previously studied CPs. The thermodynamic critical field is also evaluated using the relation, $H_C=\sqrt{\frac{H_{C1}(0)H_{C2}(0)}{{\ln} \kappa_{GL}}}$ \cite{MIB} and obtains the value, $H_C$ = 22(1) mT for Mo$_4$Re$_2$Te$_8$.

\subsubsection{Specific heat}

The bulk superconductivity is further probed by specific heat versus temperature measurement at zero applied magnetic field for Mo$_4$Re$_2$Te$_8$. \figref{Fig6:SH}(a) shows the jump in specific heat data at $T_{C,mid}$ = 3.23(8) K considering the mid-point. Above $T_C$, the normal state region of specific heat is fitted using the Debye relation, $C/T = \gamma_n + \beta_3 T^2$, where $\gamma_n$ is the Sommerfeld coefficient, and $\beta_3$ is the lattice constant. Fitting of the data provides the parameters as $\gamma_n$ = 24.7(8) mJmol$^{-1}$K$^{-2}$ and $\beta_3$ = 5.95(4) mJmol$^{-1}$ K$^{-4}$. Further, the change in superconducting transition temperature with applied magnetic field is shown in \figref{Fig6:SH}(b).

Specific heat measurement also give access to calculate several other parameters. Lattice constant parameter, $\beta_3$ is used to extract the information about the phonons and provides the Debye temperature $\theta_D$. Using the relation, $\theta_D = (\frac{12\pi^4 R N}{5\beta_3})^{1/3}$, where $R$ is the universal gas constant, and $N$  = 14 the number of atoms per formula unit, the estimated value is 166(1) K. The obtained value of $\theta_D$ from specific heat matches with the value from resistivity measurement. The Sommerfeld coefficient, $\gamma_n$ is related to single particle density of states at the Fermi level by expression, $\gamma_n = (\frac{\pi^2 k_B^2}{3}) D_c (E_{\mathrm{F}})$ and the operation yields $D_c(E_{\mathrm{F}})$ = 10.5(3) states  eV$^{-1}$ f.u.$^{-1}$. The information on electron-phonon coupling strength can also be extracted from McMillian theory, where the dimensionless quantity, $\lambda_{e-ph}$ is stated by the relation \cite{mcm},

\begin{equation}
\lambda_{e-ph} = \frac{1.04+\mu^{*}\mathrm{ln}(\theta_{D}/1.45T_{C})}{(1-0.62\mu^{*})\mathrm{ln}(\theta_{D}/1.45T_{C})-1.04 }
\label{eqn7:Lambda}
\end{equation}

here $\mu^*$ is typically assumed to be 0.13 for intermetallic superconductors. The values $\theta_D$ = 166(1) K and $T_C$ = 3.23(8) K, assort the value $\lambda_{e-ph}$ = 0.67(2), suggest that the telluride chevrel phase Mo$_4$Re$_2$Te$_8$ is a moderately-coupled superconductor. Even though, the high DOS at the Fermi level, large contribution to the lattice specific heat, $\beta_3$ and parameter $\gamma_n/T_C$ = 7.6 mJmol$^{-1}$ K$^{-3}$ of  Mo$_4$Re$_2$Te$_8$ are comparable to SnMo$_6$S$_8$, PbMo$_6$S$_8$ and Ag$_{1-x}$Mo$_6$S$_8$, possessing strong electron-phonon coupling, Mo$_4$Re$_2$Te$_8$ has moderate coupling. The ratio $\gamma_n/T_C$ indicates the scale of transition temperature with the density of states at $E_{\mathrm{F}}$ \cite{2gsc,mbsc}.

The symmetry associated with the superconducting gap can be understood by analyzing the electronic specific heat temperature dependence in the superconducting region. The electronic contribution to the specific heat is calculated by subtracting the phononic contribution from the total specific heat at zero field. While theoretically can be computed from $C_{el} = t \frac{dS}{dt}$, with $S$ being the entropy in the superconducting region and within BCS approximation has the form,

\begin{equation}
S = -\frac{6\gamma_n}{\pi^2}\left(\frac{\Delta(0)}{k_{B}}\right)\int_{0}^{\infty}[ \textit{f}\ln(f)+(1-f)\ln(1-f)]dy 
\label{eqn8:BCS1}
\end{equation}

where $f(\xi) = [ \exp (E(\xi))/k_BT)+1]^{-1}$ is the Fermi function. $E(\xi) = \sqrt{\xi^2 + \Delta(t)^2}$ is the excitation energy of the quasiparticle measured relative to the Fermi level, with $y = \xi/\Delta(0)$ and $\Delta(t)$ is the temperature-dependent gap function. In the isotropic $s$-wave BCS approximation, gap function can be written as, $\Delta(t)$ = tanh$\{1.82(1.018[(\mathit{1/t}) - 1])\}^{0.51}$ with $ t = T/T_C $. \figref{Fig6:SH}(c) displays the electronic specific heat, $C_{el}$ variation with temperature, fitted with the BCS $s$-wave model. The well fitted data in the superconducting region yields $\Delta(0)/k_B T_C$ = 1.81(4) while the specific heat jump value $\Delta C_{el}/\gamma_n T_C$ = 1.44(8). Both these values, $\Delta(0)/k_B T_C$ and $\Delta C_{el}/\gamma_n T_C$ are close to the BCS predicted value (1.76 and 1.43) in a weak coupling limit. Hence, indicates  Mo$_{4}$Re$_{2}$Te$_{8}$ as a weakly coupled superconductor. However, to know the exact structure of the superconducting gap and the nature of the pairing state, low-temperature specific heat below 1.9 K is required.

\subsubsection{Electronic properties and the Uemura plot}

\begin{figure}
\includegraphics[width=1.0\columnwidth, origin=b]{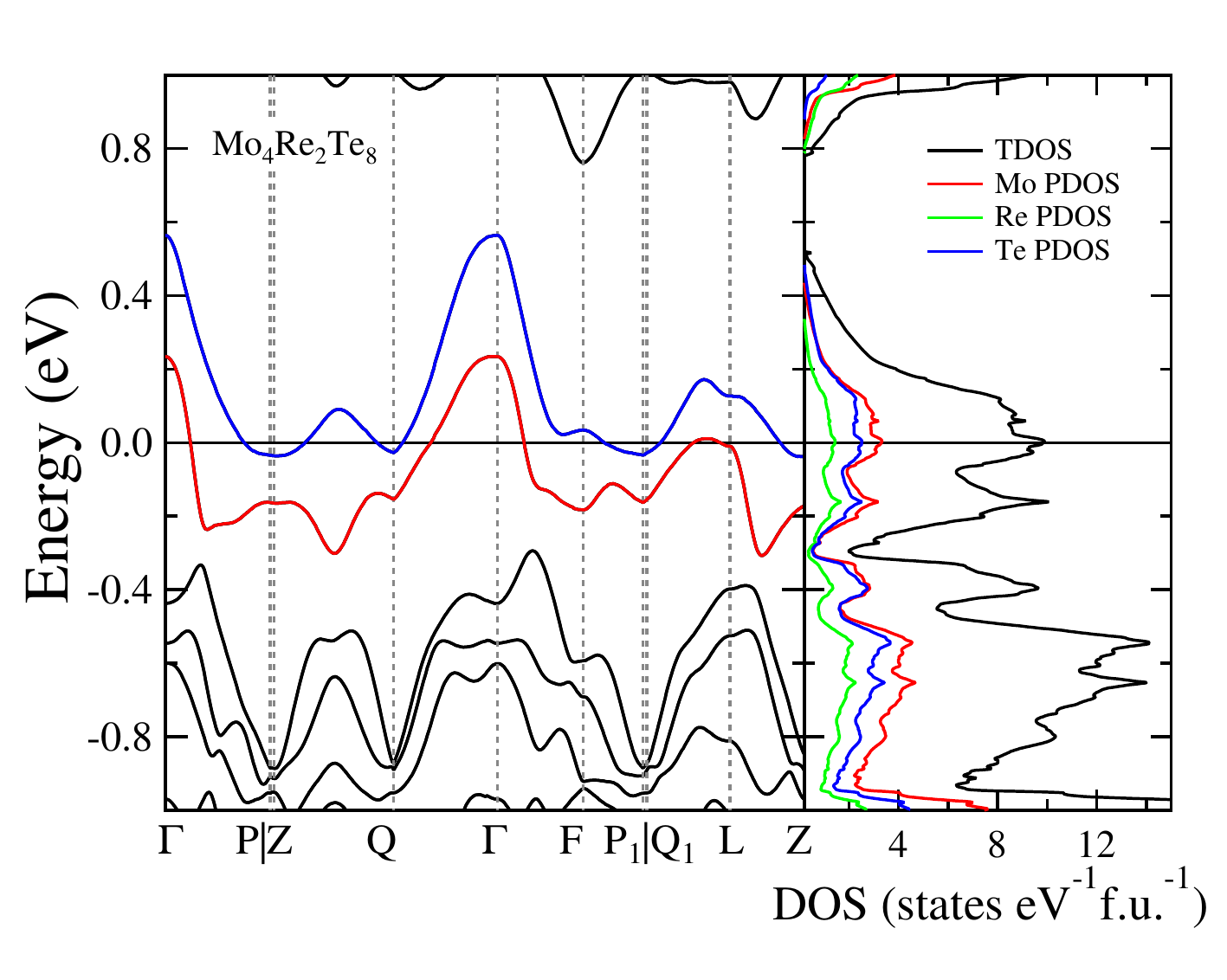}
\caption{\label{Fig5:bands} Band dispersion and density of states of Mo$_{4}$Re$_{2}$Te$_{8}$ calculated with GGA+SO.} \end{figure}

To get an insight about electronic properties, we have performed band structure calculations using the full potential linear augmented plane wave (FP-LAPW) method as implemented in Wien2k \cite{wien2k1, wien2k2}. We used $k$ point mesh of 10 $\times$ 10 $\times$ 10 within the first Brillouin zone, and the generalized gradient approximation (GGA) exchange-correlation functional of Perdew, Burke and Ernzerhof (PBE) \cite{PBE}. SOC was included for all the elements in the calculations. The energy and charge convergence criteria was set to 0.01 meV and 10$^{-4}$ electronic charge per f.u., respectively. The Mo$_{4}$Re$_{2}$Te$_{8}$ structure was simulated by substituting Re at the place of two Mo atoms in Mo$_{6}$Te$_{8}$. The electronic band structure between high symmetry points \cite{bz} in the Brillouin zone and DOS of Mo$_{4}$Re$_{2}$Te$_{8}$ are shown in \figref{Fig5:bands}. The inclusion of SOC for Re significantly modifies the band dispersion while SOC in Mo and Te do not have much influence, indicating high SOC of Re plays an important role in Mo$_{4}$Re$_{2}$Te$_{8}$. Two doubly degenerate bands, B$_1$ and B$_2$ shown in red and blue colour respectively, are dispersing in large energy window appear in the vicinity of Fermi level ($E_{\mathrm{F}}$) and crosses $E_{\mathrm{F}}$ at different $k$ points of the Brillouin zone. Fermi level lies almost at the middle of band B$_1$ while band B$_2$ remains almost empty. The system is close to cubic (Rhombohedral angle = 92.12$^{\degree}$) and the high symmetry points P, Z, Q, P$_1$ and Q$_1$ are almost equivalent where a small electron pocket (band B$_2$) is seen in the band structure. At the same time, a large hole pocket is centred around $\Gamma$ point (band B$_1$) as also observed from the Hall measurement suggesting dominant hole carriers.

Strong hybridization between Mo/Re $d$ states and Te $p$ states is observed and seen in the DOS plot. The calculated value of total DOS at $E_{\mathrm{F}}$, is 9.87 states eV$^{-1}$f.u.$^{-1}$ in qualitative agreement with values obtained from the specific heat experiment.

\begin{figure}
\includegraphics[width=1.0\columnwidth]{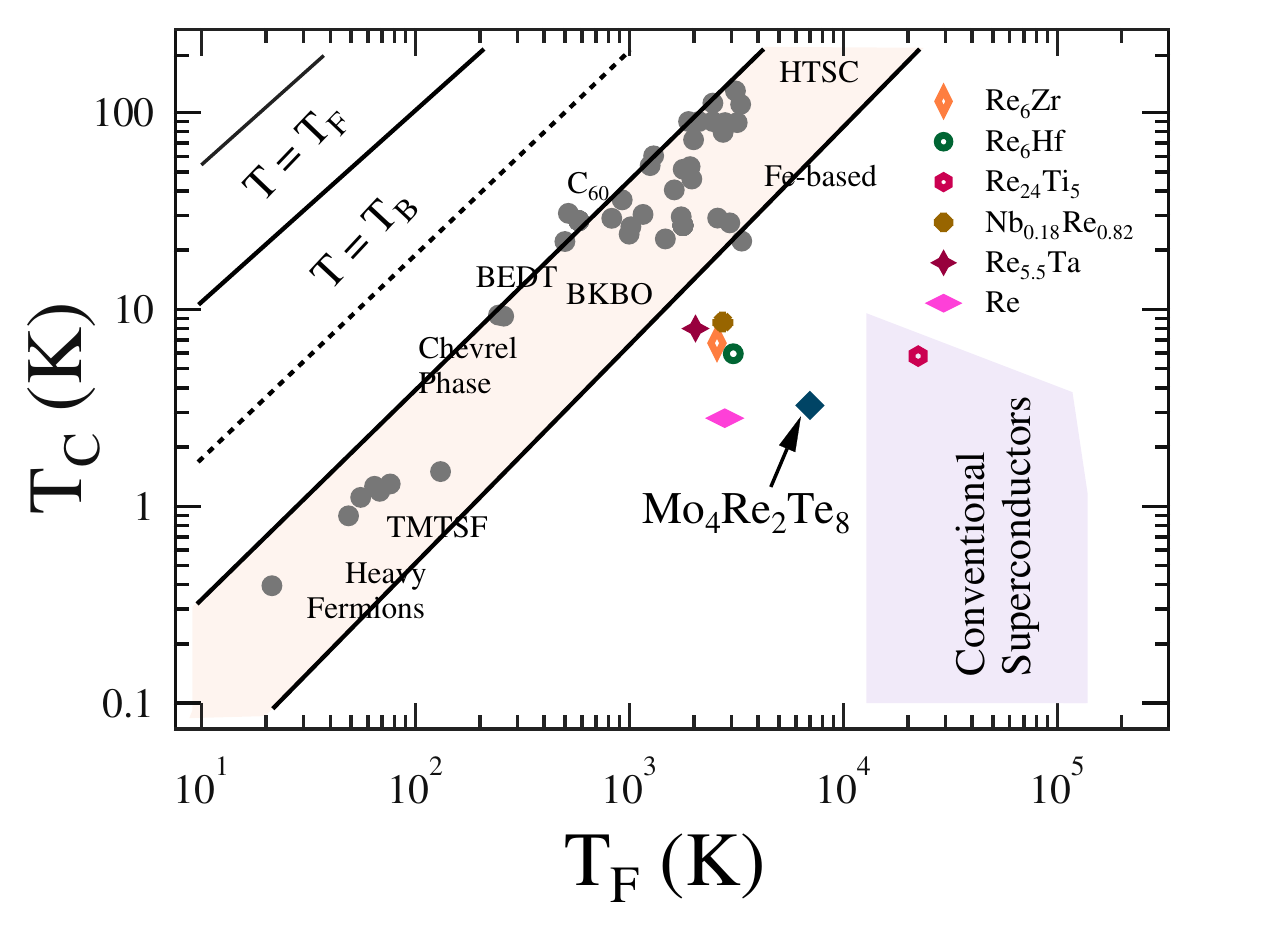}
\caption{\label{Fig7:UP} A plot between the $T_{C}$ and the $T_{\mathrm{F}}$ representing the unconventional superconductors family. Mo$_4$Re$_2$Te$_8$ is shown, which is close to conventional superconductors band region  regardless of the other Chevrel phase  which lies in the unconventional band \cite{Unconv_1,Unconv_2}. } 
\end{figure}

Further, in order to quantify the London penetration depth, $\lambda_L$, electronic mean free path, $l_e$ and to verify the dirty limit superconductivity for the Mo$_4$Re$_2$Te$_8$ a set of equations are implemented. The Fermi wave vector, $k_{\mathrm{F}}$ is estimated from quasiparticle number density by the relation, $k_{\mathrm{F}} = (3 \pi^2 n)^{1/3}$, here $n = 1.21(3) \times 10^{28}$ m$^{-3}$ from normal Hall measurement, hence $k_{\mathrm{F}}$ = 0.71(2) $\text{\AA}^{-1}$. Sommerfeld coefficient, $\gamma_n$ and $k_{\mathrm{F}}$ is used for effective mass value estimation by the relation, $m^{*} = (\hbar k_{\mathrm{F}})^2 \gamma_n/\pi^2 n k_B $, with $k_B$ being the Boltzmann constant, $m^{*}$ becomes 3.2(3) $m_e$. Moreover, in consideration of Drude's model, the mean free path is defined as, $l_e$ = $v_{\mathrm{F}} \tau$, where $\tau$ is the scattering time and given by $\tau^{-1}= n e^2 \rho_0/m^*$. $v_{\mathrm{F}}$ is the Fermi velocity stated as $v_{\mathrm{F}} = \hbar k_{\mathrm{F}}/m^*$. Including the respective values of $m^{*}$, $k_{\mathrm{F}}$, $n$, and residual resistivity, $\rho_0$ = 2.95(1) m$\ohm$-cm, the scattering time, $\tau$ and the Fermi velocity $v_{\mathrm{F}}$ are evaluated, which provides mean free path, $l_e$ = 0.81(9) \text{\AA}. From BCS theory \cite{Coh_Leng}, the coherence length $\xi_0$ is approximated as $0.18 \hbar v_{\mathrm{F}}/k_B T_C$, hence becoming 1074(202) $\text{\AA}$ for $T_C$ = 3.23 K. The huge difference between the coherence length and mean free path, $\xi_0 \gg$ $l_e$ puts Mo$_4$Re$_2$Te$_8$ in the dirty limit superconductor. Further, London penetration depth $\lambda_{L}$, is expressed as, $\lambda_{L}=\left(\frac{m^{*}}{\mu_{0}n e^{2}}\right)^{1/2}$ and calculated to be 86(1) nm. The characteristics nature of poor metal with the short mean free path of the known CPs is also observed in our telluride CP compound Mo$_4$Re$_2$Te$_8$.

Chevrel phase has been classified as unconventional superconductor based according to their $\frac{T_{C}}{T_{\mathrm{F}}}$ ratio provided by Uemura et al. \cite{Uemura}. The ratio for Chevrel phase lies in the range 0.01 $\leq$ $\frac{T_{C}}{T_{\mathrm{F}}}$ $\leq$ 0.1 with high-$T_C$, organic, heavy-fermion, and other unconventional superconductors. For 3D system, assuming spherical Fermi surface, the Fermi temperature, $T_{\mathrm{F}}$ is given by the relation \cite{Tf}

\begin{equation}
 k_{B}T_{{\mathrm{F}}} = \frac{\hbar^{2}}{2}(3\pi^{2})^{2/3}\frac{n^{2/3}}{m^{*}}, 
\label{eqn9:tf}
\end{equation}

where $n$ is the quasiparticle number density per unit volume, and $m^{*}$ is the effective mass of quasiparticles. Considering the values of $n$ and $m^{*}$ listed in Table II, from equation \ref{eqn9:tf}, we get $T_{{\mathrm{F}}}$ = 6980(845) K for Mo$_4$Re$_2$Te$_8$. The ratio  $\frac{T_{C}}{T_{{\mathrm{F}}}}$ = 0.0005 placing Mo$_4$Re$_2$Te$_8$ in between the region of Re based compounds having Re$_{24}$Ti$_{5}$ \cite{r24t5} on its left and other compounds on its right side of same composition \cite{re6hf,re6zr,Re,re5ta}, with the Re element \cite{Re}, as shown in \figref{Fig7:UP} .

\begin{table}[h!]
\caption{Parameters in the superconducting and normal state of Mo$_4$Re$_2$Te$_8$}
\begingroup
\setlength{\tabcolsep}{12pt}
\begin{tabular}{c c c} 
\hline\hline
Parameters & Unit & Mo$_4$Re$_2$Te$_8$ \\ [1ex]
\hline
$T_{C}$& K&  3.26(3)\\             
$H_{C1}(0)$& mT& 0.49(1)\\               
$H_{C2}^{\rho}(0)$& T&  5.78(4)\\
$H_{C2}^{P}(0)$& T&  5.99(5)\\
$H_{C2}^{Orb}(0)$& T&  3.55(4)\\
$\xi_{GL}(0)$& $\text{\AA}$ &  80(1)\\
$\lambda_{GL}(0)$& nm & 1325(40)\\
$k_{GL}$ & &  165(7)\\
$\gamma_{n}$&  mJ mol$^{-1}$ K$^{-2}$& 24.7(8)\\
$\theta_{D}$& K& 166(1)\\
$\Delta C_{el}/\gamma_{n}T_{C}$& & 1.44(8)\\
$\Delta(0)/k_{B}T_{C}$&   & 1.81(4)\\ 
$v_{{\mathrm{F}}}$& 10$^{5} $m s$^{-1}$&  2.6(2)\\
$n$& 10$^{28}$m$^{-3}$&  1.21(3)\\
$T_{{\mathrm{F}}}$& K & 6980(845)\\
$T_{C}/T_{{\mathrm{F}}}$& & 0.0005(1)\\
$m^{*}$/$m_{e}$&   & 3.2(3)\\
[1ex]
\hline\hline
\end{tabular}
\endgroup
\end{table} 

Irrespective of short mean free path, high DOS at Fermi level with moderate electron-phonon coupling in pseudobinary compound Mo$_4$Re$_2$Te$_8$, the exotic character of the extremely high upper critical field value of CPs is absent. As these parameters but with the strong electron-phonon coupling were accounted to be the possible reason for high upper critical field value in well-known CPs \cite{bC1,mc}. Hence, strong electron-phonon coupling might be a certain possible requirement to enhance the upper critical field beyond a limit in CPs. However, the comparable value of the upper critical field to the Pauli paramagnetic limit also suggests an unconventional nature in its superconducting ground state. Notably, such a relatively large value of the upper critical field has not been observed in any other Te superconductors, excluding the iron-based chalcogenide superconductor \cite{fst}. Further, the Uemura classification of Mo$_4$Re$_2$Te$_8$ places it in an interesting position between the two different Re concentration systems. Hence, more experimental details are required to elucidate the associated pairing mechanism with the superconducting ground state in this Chevrel phase, and muon spectroscopy is required to study the time-reversal symmetry.

\section{CONCLUSION}

To summarize, pseudobinary telluride Chevrel phase Mo$_4$Re$_2$Te$_8$ is investigated, and the superconducting transition at $T_C$ = 3.26(3) K is recorded. XRD analysis confirms the phase purity and trigonal structure crystallization. AC resistivity shows the poor metallic character of the sample with a short mean free path. The magnetization measurements suggest the type-II superconductivity with an upper critical field, $H_{C2}(0)$ = 5.78(4) T, close to the Pauli limiting field. The specific heat data measured in zero applied magnetic field presents the jump value $\Delta C_{el}/\gamma_n T_C$ = 1.44(8) with fully gapped superconductivity in Mo$_4$Re$_2$Te$_8$. Initial band structure calculations also depict significant contribution of Re in DOS at the Fermi level with notable spin-orbit coupling impact of Re in band dispersion. A Detailed microscopic study (e.g. muon spectroscopy) combined with detailed theoretical calculations of electronic structure is required to understand the superconducting ground state and the role of the Re on the superconducting properties of this compound.

\section{Acknowledgments} A. Kataria acknowledges the funding agency Council of Scientific and Industrial Research (CSIR), Government of India, for providing SRF fellowship (Award No: 09/1020(0172)/2019-EMR-I). R.~P.~S.\ acknowledges Science and Engineering Research Board, Government of India for the Core Research Grant CRG/2019/001028.

\end{document}